# Functionalized Graphene for High Performance Two-dimensional Spintronics Devices

Linze Li,[†] Rui Qin,[†] Hong Li, Lili Yu, Qihang Liu, Guangfu Luo, Jing Lu,[*] and Zhengxiang Gao

State Key Laboratory of Mesoscopic Physics and Department of Physics,

Peking University, Beijing 100871, P. R. China

[†]These authors contributed equally to this work.

*Corresponding author: jinglu@pku.edu.cn

*Abstract:* Using first-principles calculations, we explore the possibility of functionalized graphene as high performance two-dimensional spintronics device. Graphene functionalized with O on one side and H on the other side in the chair conformation is found to be a ferromagnetic metal with a spin-filter efficiency up to 85% at finite bias. The ground state of graphene semi-functionalized with F in the chair conformation is an antiferromagnetic semiconductor, and we construct a magnetoresistive device from it by introducing a magnetic field to stabilize its ferromagnetic metallic state. The resulting room-temperature magnetoresistance is up to 5400%, which is one order of magnitude larger than the available experimental values.

1. **Introduction**

Graphene has a long spin relaxation time and length[1-6] due to a small spin-orbit coupling of carbon atoms, and this makes graphene a promising material in applications of spintronics. Spin-valve and spin filter are two kinds of popular spintronics devices. Graphene-based spin-valves have been experimentally constructed, but the resulting magnetoresistance (MR) is quite small. A 10% MR is observed in a spin-valve with a graphene wire contacted by two soft magnetic electrodes at 300 K.[7] A spin-valve consisting of a graphene flake and ferromagnetic electrodes shows a 12% MR at 7 K when a MgO tunnel barrier is inserted at the graphene/electrode interface.[8] Unlike graphene, which is a nonmagnetic zero-bandgap semiconductor, zigzag graphene nanoribbons (ZGNRs) have magnetic moment on the two edges. High performance spin-valve with giant magnetoresistance (GMR) can be constructed by either using a ferromagnetic ZGNR connected to two ferromagnetic electrodes[9-10] (first type) or using an antiferromagnetic ZGNR connected to two metal electrodes[11] (second type). The first type device functions via changing the relative direction of the local magnetic field applied on the electrodes and the second type device does by applying a magnetic field on the antiferromagnetic ZGNR. ZGNRs are also predicted to be a half-metal when a transverse electrical field is applied,[12] the two edges are differently functionalized,[13] or they are rolled into nanoscrolls.[14] However, at present graphene nanoribbons with nanometer scale width cannot be produced with desirable experimental control and production of dense arrays of ordered graphene nanoribbons remains a big challenge.[15-16] Moreover, carrier mobility in ultra-narrow graphene nanoribbons is usually not as high as that in large area graphene.[17-18] One fundamental question arises: Is it possible to fabricate high performance spin-valve and spin filter from two-dimensional graphene instead of one-dimensional ZGNRs?

Functionalization of graphene is a possible scheme to attain such a goal. Fully hydrogenated graphene, which is referred to as "graphane", was predicted theoretically[19] and later synthesized through two different chemical approaches.[20-21] Graphene functionalized by groups beyond the hydrogen such as graphene oxide[22] and graphene fluoride[23] have also been synthesized. Graphane is a nonmagnetic semiconductor with a direct bandgap of 3.43 eV,[24] and graphene fluoride produced by a complete fluorination of graphene is predicted to have a band gap of ~3.5 eV. A computational work using density functional theory (DFT) suggested that when the hydrogen atoms on one side of the graphane are removed, the resulting semi-hydrogenated graphene in the chair conformation, which is referred to as "graphone", is a ferromagnetic semiconductor with an indirect bandgap of 0.46 eV.[25] The cause lies in that half-hydrogenation makes the electrons in the unhydrogenated carbon atoms localized and unpaired and the magnetic moments at these sites couple ferromagnetically. In this article, using the DFT and nonequilibrium Green's function (NEGF) method, we explore the chance to fabricate high performance spintronics devices from functionalized graphenes. We consider two types of functionalization schemes. One is functionalization on one side of graphene, and the other is different functionalizations on the two sides of graphene. Highly polarized metallic ferromagnet is obtained when graphene is functionalized with O on one side and H on the other side in the chair conformation, and GMR is obtained in a spin-valve based on graphene functionalized by F on one side in the chair conformation. New avenues are therefore opened for application of graphene in high performance two-dimensional spintronics devices.

2. Model and Method

We consider five different functionalized graphenes: graphene semi-functionalized with F (F-graphene), O (O-graphene), or OH (OH-graphene) on only one side and graphene fully-functionalized with H on one side and with F (F-graphene-H) or O (O-graphene-H) on the other side. As each functionalized graphene has chair and boat two conformations, we totally calculate ten different structures. We have constructed supercells consisting of 8 carbon atoms to check the magnetism of our functionalized graphenes. All supercells are large enough to ensure that the vacuum space is at least 10 Å, so that the interaction between functionalized graphenes and their periodic images can be safely avoided.

The geometry optimization and electronic properties are calculated by using ultrasoft pseudopotentials plane-wave method, as implemented in the CASTEP code.[26-27] The generalized gradient approximation (GGA) of the Perdew–Burke–Ernzerhof (PBE)[28] form is employed for the exchange–correlation functional. The reciprocal space was represented by Monkhorst-Pack[29] special $k$-point scheme with $12 \times 12 \times 1$ grid meshes. The geometrical structures are relaxed without any symmetry constraints with a plane-wave cutoff energy of 400 eV. The convergence of energy and force are set to $1 \times 10^{-5}$ eV and 0.03 eV/Å, respectively. The accuracy of our procedure was tested by calculating the C-C bond length of graphene: our calculated result of 1.42 Å is the same as the experimental value.

Two-probe model is constructed to study the transport properties. The transport properties are computed by using the DFT coupled with the NEGF formalism implemented in the ATK code.[30-32] The local density approximation (LDA) and norm-conserving pseudopotentials of the Troullier-Martins type[33] are used. Single-ζ basis set is used and the mesh cutoff is chosen as 150 Ry, and the electron temperature is set to 300 K. The structures of the scattering region are optimized until the maximum atomic forces are less than 0.03 eV/Å. The spin-resolved current $I_\sigma$ under bias voltage $V_{bias}$ is calculated with the Landauer-Büttiker formula [34]:

$$I_\sigma(V_{bias}) = \frac{e}{h} \int \{T_\sigma(E,V_{bias})[f_L(E,V_{bias}) - f_R(E,V_{bias})]\}dE \tag{1}$$

where $T_\sigma(E,V_{bias})$ is the spin-resolved transmission probability, $f_{L/R}(E,V_{bias})$ is the Fermi-Dirac distribution function for the left (L)/right (R) electrode, and σ is a spin index.

### 3. Results and Discussion

We begin our study by optimizing the geometric structures of the ten different functionalized graphenes in their nonmagnetic state. The difference in geometry between different magnetic states is negligibly small. We present the chair and boat conformations of the functionalized graphenes in Figure 1 and Figure 2, respectively. In all the structures except the O-graphene, C atoms of graphene layer are corrugated, forming two atom sub-layers, and the F and O atoms are above one C atom. We display the optimized structure of the chairlike and boatlike F-graphene-H in Figure 1 (a) and Figure 2 (a), respectively. Structures of the O-graphene-H are similar to them. In the structure of the OH-graphene (Figure 1 (b) and 2 (b)), the H atom tends to site above the center of the hexagonal ring of graphene. As shown in Figure 1 (c) and 2 (c), the O atoms of the O-graphene site above the carbon-carbon bonds, form two bonds with two carbon atoms ([2+1] cycloaddition), and thus keep all C atoms staying in one layer. All boat conformations has three different types of C-C bonds (A, B, and C types in Figure 2) while chair conformations except the chairlike O-graphene (A and B types in Figure 1 (c)) have only one. The geometric parameters of all functionalized graphenes are listed in Table 1, in comparison with the parameters of the graphone and graphane. The C-C bond lengths inside the graphene of all the chair conformations (from 1.50 to 1.56 Å) are larger than that of the pristine graphene (1.42 Å), while in the boat conformations the bond lengths between two unfunctionalized C atoms (from 1.35 to 1.38 Å) are less than it. The thicknesses of the corrugation of the fully-functionalized graphenes (chair: from 0.46 to 0.54 Å, boat: from 0.59 to 0.65 Å) are larger than those of the semi-functionalized graphenes (chair: from 0.30 to 0.35 Å, boat: from 0.40 to 0.43 Å), and the thickness of the corrugation of the boat conformation of each functionalized graphene is larger than that of the chair one.

We compute the binding energy per group to examine the stability of all the functionalized graphenes. Here, the binding energy per group, $E_b$, is defined as:

$$E_b = (E_G + m \times E_{fg} - E_{FG})/m \qquad (2)$$

where $E_G$, $E_{fg}$, and $E_{FG}$ are the respective energies of the pristine graphene, a functional group, and the functionalized graphene, and $m$ is the number of the attached groups. If graphene is functionalized with two different kinds of groups (group(1) and group(2)), $m$ should be replaced by $m(1) + m(2)$, and $m \times E_{fg}$ should be replaced by $m(1) \times E_{fg}(1) + m(2) \times E_{fg}(2)$. The stability of different structures can be evaluated by binding energies: those with larger binding energies are more thermodynamically stable.

Table 2 presents the binding energies of all functionalized graphenes and those of graphane and graphone for comparison. The fully-saturated graphenes (which means that all C atoms in graphene layer are saturated, such as O-graphene, F-graphene-H, O-graphene-H, and graphane) are more stable than the semi-saturated graphenes (which means that only half C atoms in graphene layer are saturated, such as F-graphene, OH-graphene, and graphone). The O-graphene and O-graphene-H are more stable than graphane by 0.77 and 0.43 eV per group, respectively. And the stability of the F-graphene-H is similar to that of graphane. The F-graphene is as stable as the graphone, while the OH-graphene is the most unstable. Chair conformations of the fully-saturated graphenes are slightly more stable than the boat ones by 0.06 ~ 0.15 eV per group, whereas boat conformations of the semi-saturated graphenes are much more stable than the chair ones by 0.63 ~ 0.88 eV per group. Therefore, both synthesized graphane[20-21] and graphene fluoride[23] should be in the chair conformation. In the following work, the less stable boatlike fully-saturated graphenes will not be considered. As for semi-saturated graphenes, the conformation depends on the actual reaction path. If we solely functionalize one side of the graphene with the other side intact, we will get the more stable the boat conformation. If we first get two-side functionalized graphenes and then remove the functional groups on one side, we are highly likely to get the chair conformation because it is difficult to reconstruct from the chair to the boat configuration due to the totally different group alignments of the two configurations. The formerly predicted ferromagnetic graphone adopted the chair conformation.[25]

Next, we study the magnetism of the functionalized graphenes. According to our results, all the boat conformations have no magnetism. Ferromagnetically (FM) coupled (Figure 3 (a)), antiferromagnetically (AF) coupled (Figure 3 (b)), and nonmagnetic (NM) three states are considered for chair conformations. Both the chairlike O-graphene and H-graphene-F turn out to be nonmagnetic since all atoms in the two structures are saturated. The relative energies of different magnetic configurations of the chairlike F-graphene, OH-graphene, and H-graphene-O are given in Table 3. Both the chairlike F-graphene and the OH-graphene have AF ground state, while the chairlike H-graphene-O has FM ground state. Using the mean field theory, Curie or Neel temperature can be estimated by energy differences between FM and AF states of chairlike functionalized graphenes (Table 3). Neel temperature of the chairlike F-graphene ($T_N$ = 754 K) and Curie temperature of the chairlike O-graphene-H ($T_C$ = 522 K) are much higher than Neel temperature of chairlike OH-graphene ($T_N$ = 29 K). The induced magnetic moments in the chairlike F-graphene and OH-graphene are mainly localized on the unfunctionalized C atoms with a value of 0.80 and 0.88 $\mu_B$ respectively (Figure 3 (c) shows the spin density of the AF chairlike F-graphene), while the magnetic moments in the chairlike H-graphene-O are chiefly localized on the O atoms ($M_O$ = 0.74 $\mu_B$) and secondarily on the unfunctionalized C atoms ($M_C$ = 0.14 $\mu_B$) (Figure 3 (d)).

The NM boatlike OH-graphene and F-graphene are semiconductors with indirect band gap of 1.63 and 2.18 eV (See Figure S1). The NM chairlike O-graphene and H-graphene-F and AF chairlike OH-graphene are semiconductors with direct band gap of 3.52, 3.18, and 0.60 eV, respectively (See Figure S2). The band structure of the FM chairlike OH-graphene is shown in Figure 4 (a). It is highly spin polarized, and the band gap is 0.40 and 4.58 eV in the minority and majority spin channels, respectively. The band structure of the FM chairlike

O-graphene-H is shown in Figure 4 (b). Both spin channels are metallic but there is a gap just 0.40 eV above the Fermi level ($E_f$) in the majority spin channel. Figures 4 (c) and (d) show the band structures of the AF and FM chairlike F-graphene, respectively. The AF (ground state) chairlike F-graphene is a semiconductor with an indirect bandgap of 1.17 eV, and the FM state show a metallic nature. The conductivity of the chairlike F-graphene can be significantly changed if a magnetic field is applied and stabilize the FM state. This suggests that a spin-valve with GMR could be built out of the chairlike F-graphene.

The two-probe model of the chairlike O-graphene-H and F-graphene based devices are depicted in Figure 5 (a) and (b), respectively. The FM chairlike O-graphene-H itself is used as metallic electrodes; thus the scattering region is identical to the electrodes (Figure 5 (a)). We chose the semi-planar non-magnetic graphene as electrodes to connect one 3-nm-wide chairlike F-graphene sheet, and an arch deformation occurs in the scattering region upon optimization (Figure 5 (b)).

The spin polarized zero-bias transmission spectra $T(E)$ of the chairlike O-graphene-H is presented in Figure 6 (a). The transmission coefficients within the bias window in the minority spin channel are apparently larger than those in the majority spin channel. We define the spin-filter efficiency at zero bias as:

$$\text{SFE} = \frac{T_{\min}(E_f) - T_{maj}(E_f)}{T_{\min}(E_f) + T_{maj}(E_f)} \quad (3)$$

where $T_{\min}(E_f)$ and $T_{\text{maj}}(E_f)$ represent the transmission coefficient of the minority and majority spin channels at the Fermi level ($E_f$), respectively. The calculated SFE at zero bias is 41%. The spin-polarized $I$-$V_{\text{bias}}$ curves of the chairlike O-graphene-H model are shown in Figure 6 (b), where $I$ is the current density for a two-dimensional device. Obviously the total current density remains dominated by the minority spin. Figure 6 (c) presents the spatially resolved local densities of states (LDOS) evaluated at $E_f$ under 0.5 V bias of the chairlike O-graphene-H model. The LDOS of the majority spin channel on both ends is sparser than that of the minority spin, a result in agreement with the lower conductivity of the majority spin than that of the minority one. We define the spin-filter efficiency (SFE) at the finite bias voltage as:

$$\text{SFE} = \frac{I_{\min} - I_{\text{maj}}}{I_{\min} + I_{\text{maj}}} \qquad (4)$$

where $I_{\min}$ and $I_{\text{maj}}$ represent minority and majority spin current density, respectively. We present the SFE versus $V_{\text{bias}}$ curve in Figure 6 (d). With the increase of the bias voltage, the current density of the minority spin increases significantly in the lower bias and tends to saturate in the higher bias but the current density of the majority spin changes slightly (Figure 6 (c)). As a result, the SFE initially increases with the bias. The SFE increases from 56% to 85% as the bias voltage increases from 0.1 to 0.5 V and slightly decreases to 80% at $V_{\text{bias}}$ = 0.6 V and then increases to 83% at $V_{\text{bias}}$ = 0.8 V. Therefore, the ferromagnetic metallic chairlike O-graphene-H performs well in producing spin polarized current.

To understand bias dependence of SFE of the homogenous chairlike O-graphene-H based device, we study the bias dependent electronic structure of the two chairlike O-graphene-H electrode and the transmission spectrums of the device and show the results in Figure 7. At zero bias, the transmission of this homogenous device is perfect, and the transmission probability is only determined by the number of states whose momentums have a component towards the transport direction. Roughly the SFE at zero bias can be estimated from the following expression

$$\text{SFE} = \frac{N(E_f)_{\min} - N(E_f)_{\text{maj}}}{N(E_f)_{\min} + N(E_f)I_{\text{maj}}} \qquad (5)$$

where $N(E_f)_{\min}$ and $N(E_f)_{\text{maj}}$ are the densities of states (DOS) of the minority and majority spins at $E_f$, respectively. The calculated spin-resolved DOS is given in Figure S3 of Supporting Information, and the estimated SFE is 21%, which has the same sign with the exact one and the magnitude is only half the exact one (41%). At zero bias, the transmission coefficient of the minority spin slightly increases with $E$, and that of the majority spin decreases rapidly with $E$ and vanishes at $E$ = 0.24 eV due to the opening of the band gap from this energy. At finite bias $V_{\text{bias}}$, $E_f$ of the right and left electrode are shifted by $V_{\text{bias}}/2$ and $-V_{\text{bias}}/2$, respectively. The transmission spectra of the minority spin is slightly depressed by $V_{\text{bias}}$. The top of the non-trivial transmission spectra of the majority spin drops with the increasing $V_{\text{bias}}$ as a result of the drop of the valence maximum of the right electrode. Consequently, a zero transmission region appears in the bias window at $V_{\text{bias}}$ = 0.2 V (Figure 7 (b)), and its range linearly increases with $V_{bias}$. At $V_{bias}$ = 0.4 V (Figure 7 (c)), the range is 0.2 eV. As a result, the current density of the majority spin changes slightly with $V_{bias}$ because the integral area of the transmission spectra of the majority spin within the bias window is changed slightly. SFE is a ratio of the two current densities and it thus increases with $V_{bias}$ in the lower bias.

The total current densities of the FM and AF solutions versus $V_{bias}$ curves of the chairlike F-graphene model are shown in Figure 8 (a). As we expect, the total current density in the FM solution is significantly greater than that in the AF solution. Figure 8 (c) and (d) present the 0.2-V-bias transmission spectra of the FM and AF solutions, respectively. The transmission coefficients within the bias window of the FM solution are much larger than those of the AF solution. This great difference of conductance between the two solutions is due to the different conducting mechanism (metallic versus tunneling), which is also reflected from the spatially resolved LDOS at $E_f$ under 0.2-eV bias (insets of Figure 8 (c) and (d)). Magnetoresistance is defined as:

$$\mathrm{MR} = \frac{I_F - I_{AF}}{I_{AF}} \tag{6}$$

where $I_F$ and $I_{AF}$ represent current density of the FM and AF solution, respectively. Then we present the room-temperature MR versus $V_{bias}$ curve in Figure 8 (b). In the range of $V_{bias}$ = 0.2 ~ 0.4 V, the MR is up to 3000~ 5400%. The unrelaxed two-probe model gives a larger maximum MR of 7000% (See Figure S4). The maximum experimental room-temperature MR values are a few hundred percent,[35-37] and our theoretical maximum room-temperature MR is thus one order of magnitude larger than the available experimental maximum values and two orders of magnitudes larger than the maximum room-temperature MR obtained on the spin-valve built out of pure graphene.[7] Our theoretical maximum room-temperature MR for the chairlike F-graphene spin-valve are much lower than the theoretical values of $10^6$ ~ $10^9$ for the first type ZNGR spin-valves[9-10] but comparable with that of the second type ZGNR spin-valve[38] since our chairlike F-graphene spin-valve has identical working mechanism with that of the second type ZGNR spin-valve.[38] When the bias voltage increases to 0.8 eV, the current density in the AF solution increases significantly and the MR decreases to 194%. The depressed MR with the increasing bias is also found in ZGNR spin-valves.[9-10, 38]

## 4. Conclusion

By first-principles calculations, we demonstrate that functionalization of a nonmagnetic graphene can lead to stable novel magnetic materials with high spin filter efficiency and giant room-temperature magnetoresistance comparable with that of graphene nanoribbons. This renders functionalized graphene a promising material for high performance two-dimensional spintronics devices. Compared with ultra-narrow graphene nanoribbon spintronics devices, functionalized graphenes allows much larger current with lower requirement in fabrication technique and are more competitive.

**Acknowledgement.** This work was supported by the NSFC (Grant Nos. 10774003, 90626223, and 20731162012), National 973 Projects (No. 2007CB936200, MOST of China), Program for New Century Excellent Talents in University of MOE of China, Fundamental Research Funds for the Central Universities, National Foundation for Fostering Talents of Basic Science (No. J0630311).

**Supporting Information available:** Band structures of the NM boatlike OH-graphene, NM boatlike F-graphene, NM chairlike O-graphene, NM chairlike F-graphene-H, and AF chairlike F-graphene; Density of states of the FM O-graphene-H; Schematic model of an unrelaxed chairlike F-graphene based spin-valve device; $I$-$V_{bias}$ curve and bias dependence of the magnetoresistances of the unrelaxed chairlike F-graphene based spin-valve device. This material is available free of charge via the Internet at http://pubs.acs.org.

**Table 1.** Geometry parameters of differently functionalized graphenes: length of the C-C bond inside the graphene ($d_{C-C}$), distance between the functional group and its nearest C atom ($d_{FG-C}$), thickness of the corrugation of graphene ($h$), and angle ($\theta$) of $\angle$C-O-C and $\angle$C-O-H in the O-graphene and OH-graphene, respectively.

| Chair conformation | F-graphene | O-graphene | OH-graphene | Graphone | F-graphene-H | O-graphene-H | Graphane |
|---|---|---|---|---|---|---|---|
| $d_{C-C}$ (Å) | 1.50 | 1.54 (A) 1.51 (B) | 1.51 | 1.50 | 1.56 | 1.56 | 1.53 |
| $d_{FG-C}$ (Å) | 1.49 | 1.43 | 1.51 | 1.15 | 1.39 (F-C) 1.11 (C-H) | 1.36 (O-C) 1.11 (C-H) | 1.11 |
| $h$ (Å) | 0.30 | 0 | 0.35 | 0.32 | 0.47 | 0.54 | 0.46 |
| $\theta$ | - | $\angle$C-O-C: 64° | $\angle$C-O-H: 105° | - | - | - | - |
| Boat conformation | F-graphene | O-graphene | OH-graphene | Graphone | F-graphene-H | O-graphene-H | Graphane |
| $d_{C-C}$ (Å) | 1.63 (A) 1.53 (B) 1.36 (C) | 1.58 (A) 1.52 (B) 1.53 (C) | 1.68 (A) 1.54 (B) 1.38 (C) | 1.55 (A) 1.50 (B) 1.35 (C) | 1.63 (A) 1.55 (B) 1.60 (C) | 1.54 (A) 1.55 (B) 1.56 (C) | 1.57 (A) 1.54 (B) |
| $d_{FG-C}$ (Å) | 1.42 | 1.42 | 1.51 | 1.13 | 1.38 (F-C) 1.10 (C-H) | 1.45 (O-C) 1.11 (C-H) | 1.10 |
| $h$ (Å) | 0.41 | 0 | 0.43 | 0.40 | 0.63 | 0.59 | 0.65 |
| $\theta$ | - | $\angle$C-O-C: 64° | $\angle$C-O-H: 107° | - | - | - | - |

**Table 2.** Binding energies per group ($E_{Cb}$ and $E_{Bb}$ for the chair conformation and boat conformation, respectively) of functionalized graphenes, graphane, and graphone.

| Structure | F-graphene | O-graphene | OH-graphene | F-graphene-H | O-graphene-H | Graphane | Graphone |
|---|---|---|---|---|---|---|---|
| $E_{Cb}$ (eV) | 1.51 | 4.34 | 0.82 | 3.48 | 4.00 | 3.57 | 1.74 |
| $E_{Bb}$ (eV) | 2.18 | 4.20 | 1.45 | 3.33 | 3.94 | 3.47 | 2.62 |

**Table 3.** Relative energies per supercell of different magnetic configuration states (ferromagnetic: $E_{FM}$, antiferromagnetic: $E_{AF}$, and nonmagnetic: $E_{NM}$) for chairlike functionalized graphenes.

| Chairlike Structure | F-graphene | OH-graphene | O-graphene-H |
|---|---|---|---|
| $E_{FM}$ (eV) | 0.26 | 0.01 | 0 |
| $E_{AF}$ (eV) | 0 | 0 | 0.18 |
| $E_{NM}$ (eV) | 0.52 | 1.01 | 0.17 |
| Curie or Neel temperature (K) | 754 | 29 | 522 |

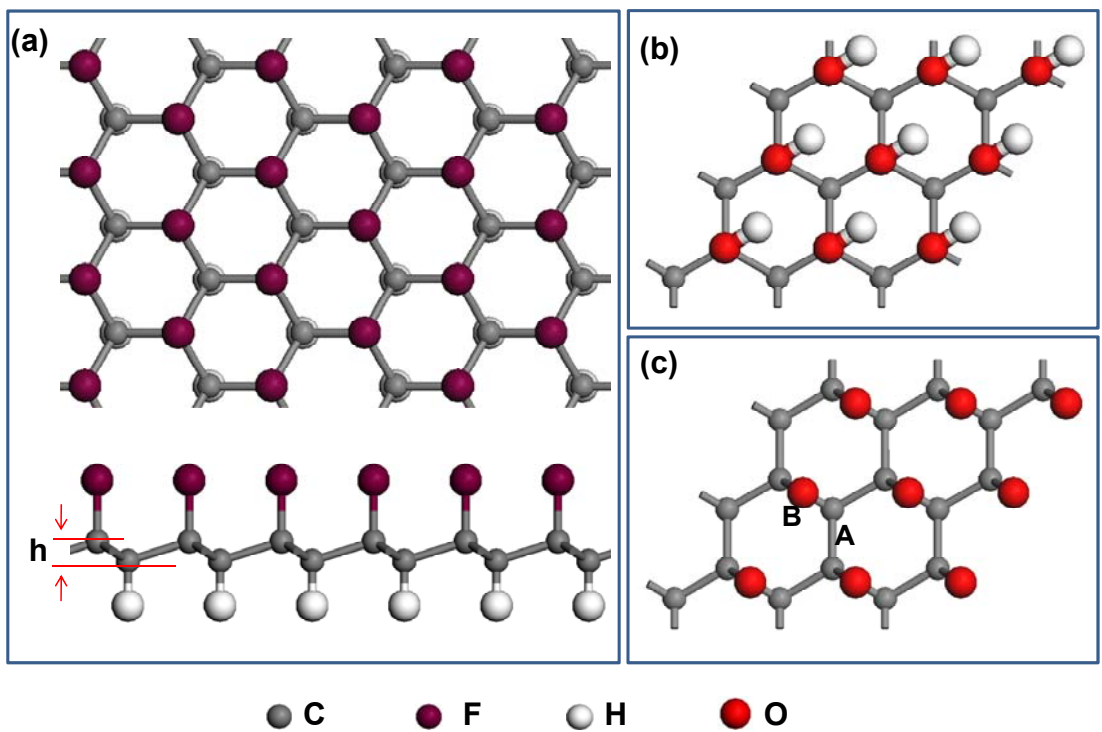

**Figure 1.** (a) Top and side views of the chairlike F-graphene-H. Top view of (b) the chairlike OH-graphene and (c) O-graphene.

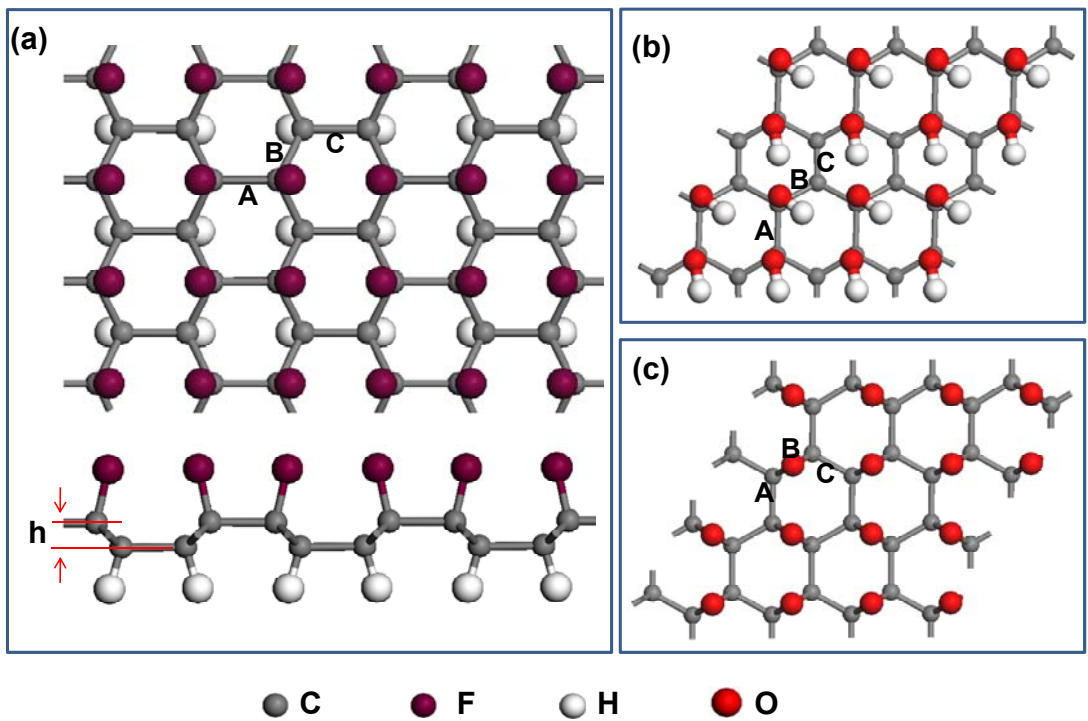

**Figure 2.** (a) Top and side views of the boatlike F-graphene-H. Top view of (b) the boatlike OH-graphene and (c) O-graphene.

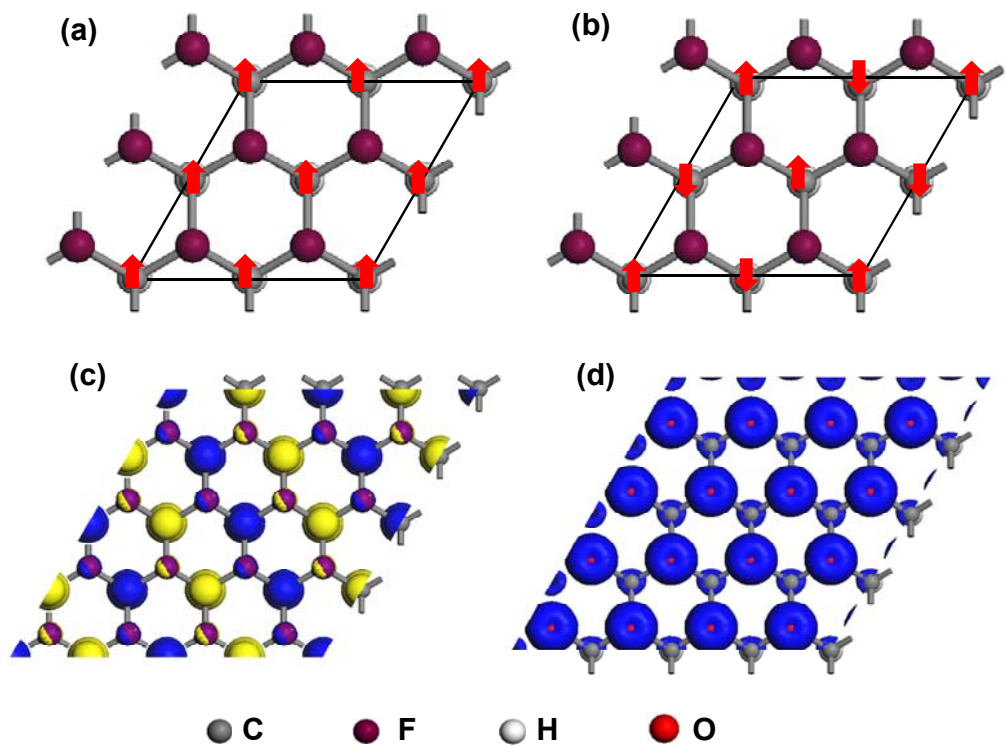

**Figure 3.** (a) Ferromagnetic and (b) antiferromagnetic configurations of chairlike functionalized graphenes. Arrows show the relative direction of magnetic moments, and rhombus marked in black shows the supercell. Spin density of (c) AF chairlike F-graphene (isovalue: 0.1 a.u.) and (d) FM chairlike O-graphene-H (isovalue: 0.08 a.u.). Blue and yellow are used to indicate the positive and negative signs of the spin, respectively.

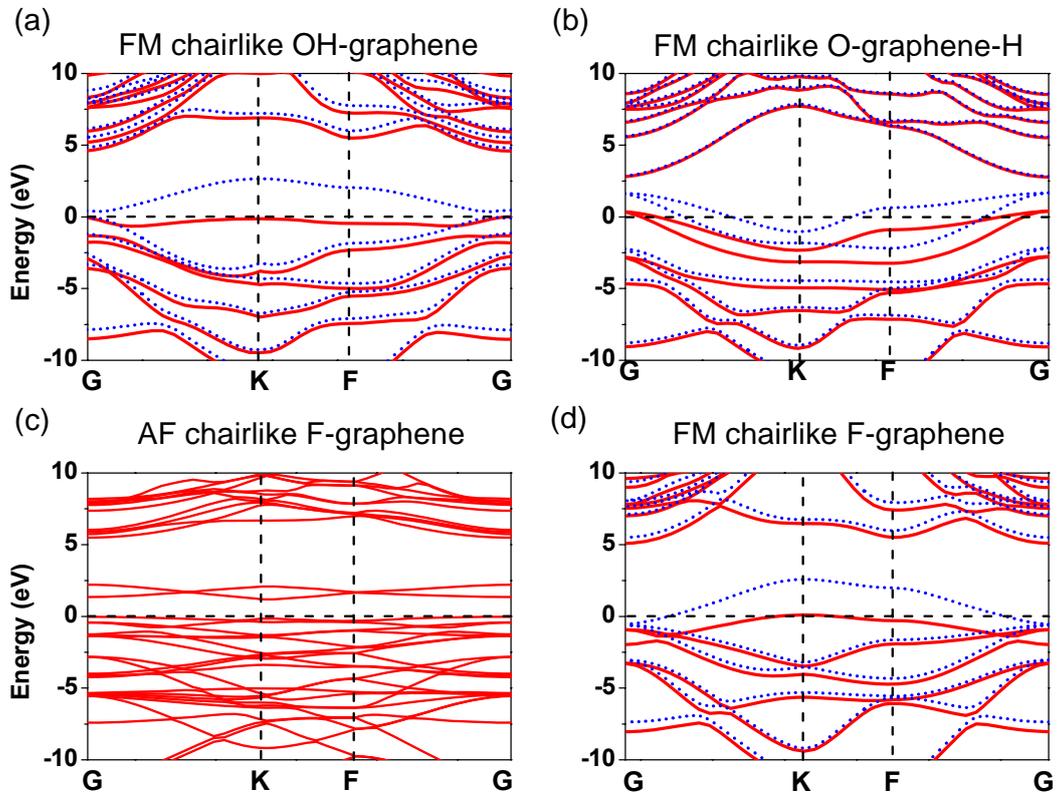

**Figure 4.** Band structures of the (a) FM chairlike OH-graphene, (b) FM chairlike O-graphene-H, (c) AF chaielike F-graphene, and (d) FM chairlike F-graphene. Red solid (blue dashed) line represents the majority (minority) spin channel. The panel (c), the two spins are degenerate. The Fermi level is set to zero.

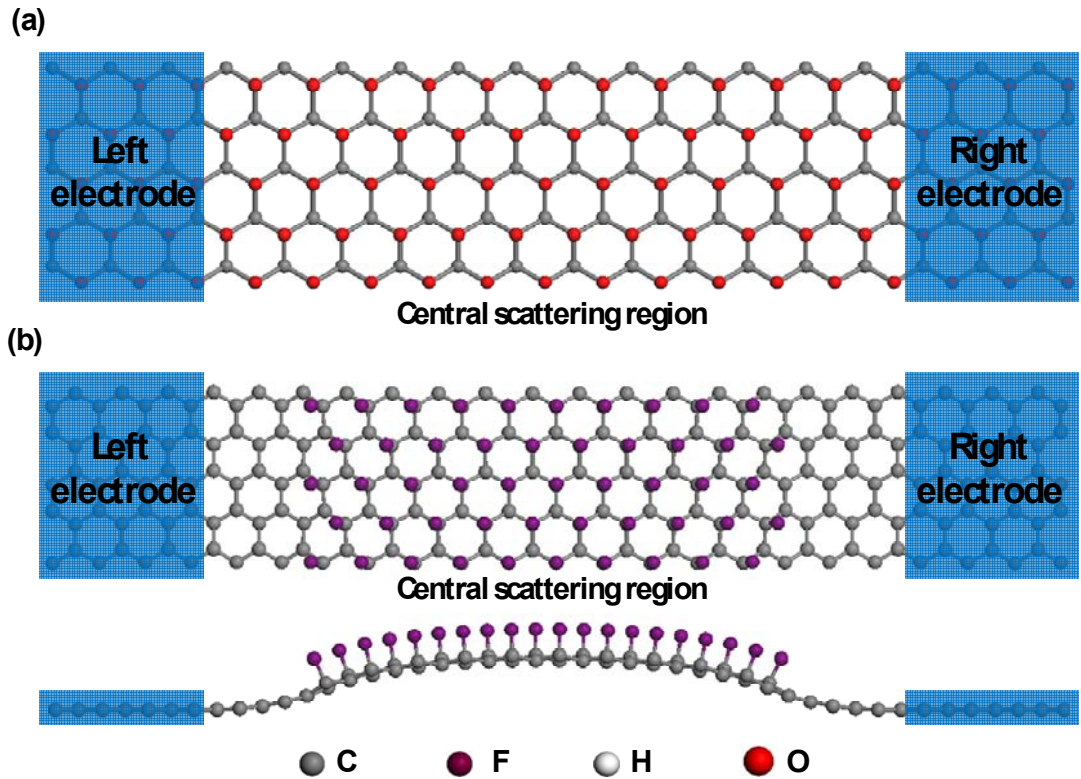

**Figure 5.** (a) Schematic relaxed two-probe model of a chairlike O-graphene-H based spin-filter device. The FM chairlike O-graphene-H itself is used as metallic electrodes. (b) Top and side views of schematic model of a chairlike F-graphene based spin-valve device. The 3-nm-wide chairlike F-graphene sheet is connected to semi-planar non-magnetic graphene electrodes.

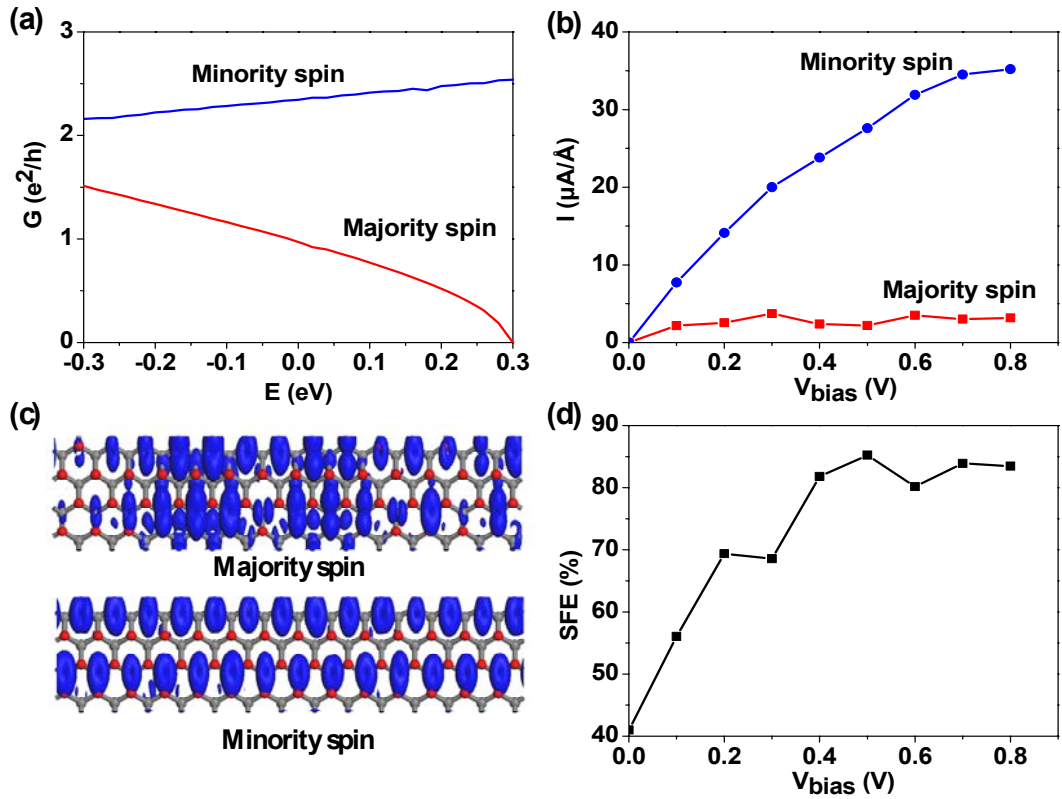

**Figure 6.** (a) Spin polarized zero-bias transmission spectra, (b) isosurfaces of the constant local densities of states evaluated at the Fermi level under zero bias (isovalue: 0.005 a.u.), (c) spin-resolved $I$-$V_{bias}$ curve, and (d) bias dependence of the spin-filter efficiency of the chairlike O-graphene-H based spin-filter device. The Fermi level is set to zero.

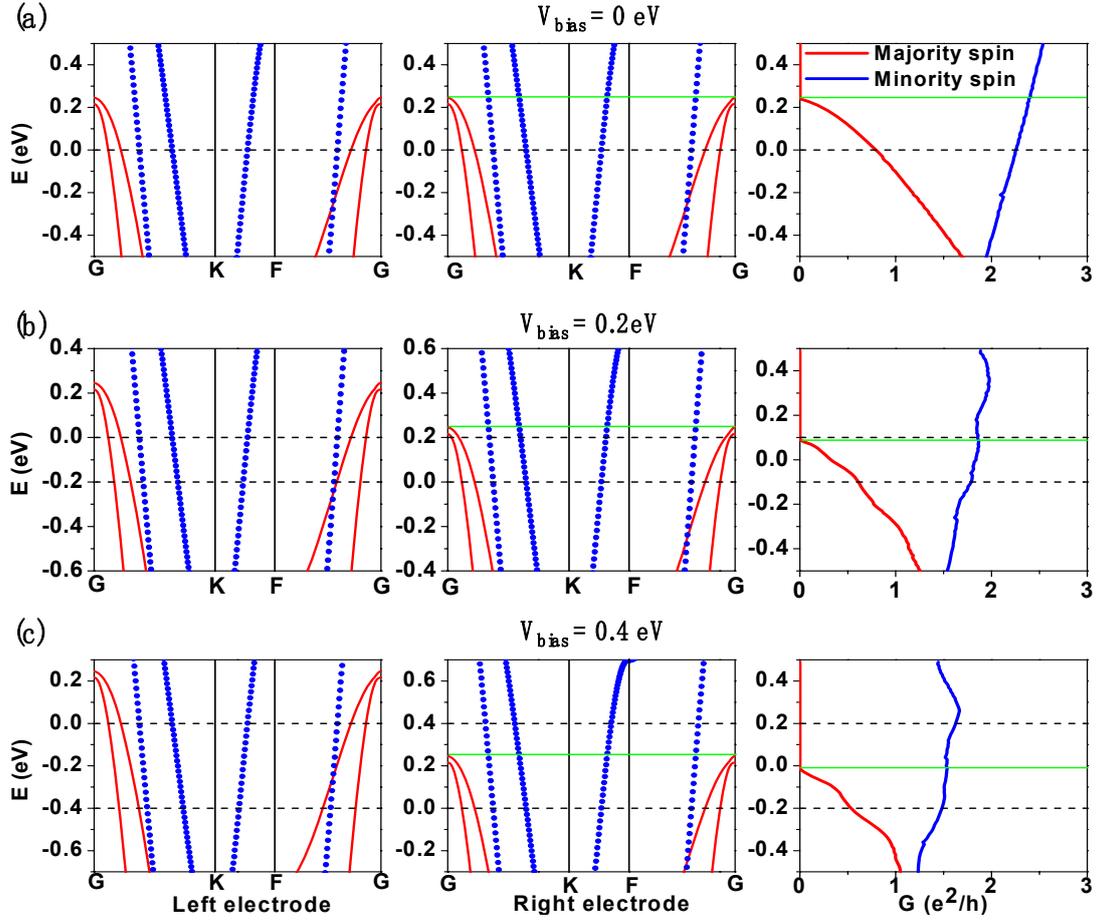

**Figure 7.** Band structures of the left (left panel) and right (middle panel) electrodes and the spin-resolved transmission spectrum (right panel) of the chairlike O-graphene-H device at a bias of (a) 0 V, (b) 0.2 V, and (c) 0.4 eV. Red solid (blue dashed) line represents the majority (minority) spin channel. The dashed black line represents the bias window. The solid green lines in the middle and right panels represent the valence maximum of the right electrode and the top of the non-trivial transmission spectra of the majority spin, respectively.

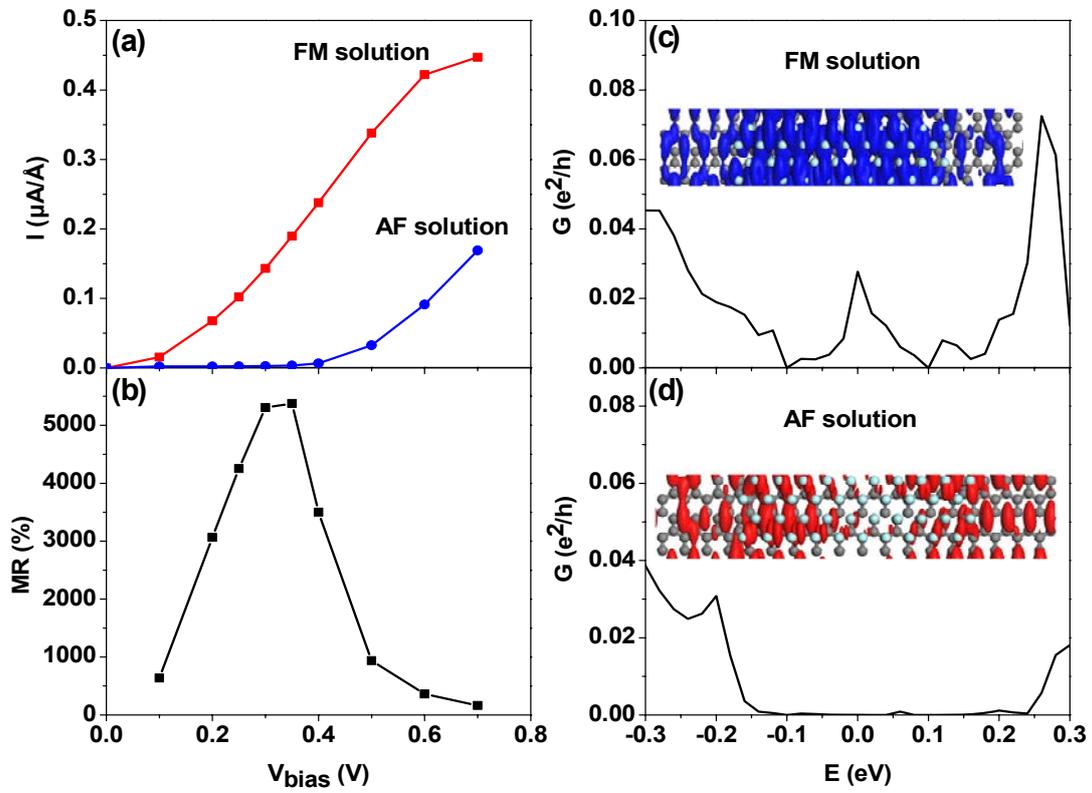

**Figure 8.** (a) $I$-$V_{bias}$ curve, (b) bias dependence of the magnetoresistances, (c,d) 0.2-V-bias transmission spectra of (c) FM solution and (d) AF solution of the chairlike F-graphene based spin-valve device. Insets: isosurfaces of the constant local densities of states evaluated at the Fermi level under 0.2-V bias (isovalue: 0.02 a.u.). The Fermi level is set to zero.

Table of contents

By first-principles calculations, we demonstrate that functionalization of a nonmagnetic graphene can lead to stable novel two-dimensional magnetic materials with high spin filter efficiency and giant magnetoresistance.

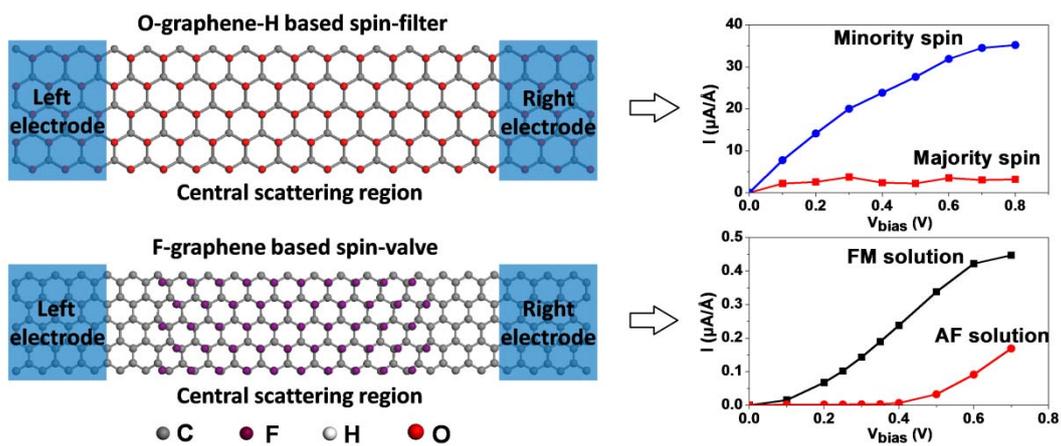